\documentclass[9pt,twocolumn,twoside]{osajnl}
\journal{ol} 
\usepackage{lineno}

\setboolean{shortarticle}{true}
\newcommand{\sq}[1]{\sqrt{\smash[b]{#1}}}
\newcommand{\p}{\partial}
\newcommand{\nn}{\nonumber}

\newcommand{\ep}{\varepsilon}

\newcommand{\vg}{\textsl{g}}
\newcommand{\om}{\omega}
\newcommand{\ta}{\theta}

\newcommand{\cH}{{\cal H}}
\newcommand{\cF}{{\cal F}}
\newcommand{\cE}{{\cal E}}

\newcommand{\wh}{\widehat}
\newcommand{\wt}{\widetilde}

\newcommand{\be}{\begin{equation}}                                       
\newcommand{\ee}{\end{equation}}
\newcommand{\ba}{\begin{eqnarray}}
\newcommand{\ea}{\end{eqnarray}}

\newcommand{\bref}[1]{(\ref{#1})}

\newcommand{\bi}[1]{\bibitem{#1}}

\newcommand{\lab}[1]{\label{#1}}

\title{Soliton blockade in bidirectional microresonators}
\author[1]{Zhiwei Fan}
\author[1,*]{Dmitry V. Skryabin}
\affil[1]{Department of Physics, University of Bath, Bath BA2 7AY, UK}
\affil[*]{Corresponding author: d.v.skryabin@bath.ac.uk}

\begin{abstract}
We report a method to control - disrupt and restore,  a   regime of the unidirectional soliton generation in a bidirectionally pumped ring microresonator. This control, i.e., the soliton blockade, is achieved by tuning pump frequency of the counterrotating field. The blockade effect is correlated with the emergence of a dark-bright nonlinear resonance of the cw-states.
\end{abstract}

\setboolean{displaycopyright}{true}

\begin{document}

\maketitle

Refractive index control with light is an important tool of achieving the functionality
of photonic circuits and optical devices, see, e.g., \cite{lip}. One of its applications is inducing shifts of the resonance frequencies 
and thereby controlling reflectivity and transmission of
resonators. Ultimate efficiency of this approach 
is achieved when  one photon interacting with one atom is enough to alter the system transmission \cite{kim,luk,tob}. A famous example of this is photon blockade \cite{kim,luk}. In this work, we  demonstrate that tuning frequency of one of the counterrotating fields in a high-Q ring microresonator can disrupt and restore  soliton transmission in the other. We term this phenomenon - soliton blockade.

Frequency conversion and soliton generation in high-Q microresonators impact many areas of modern
photonics from precision spectroscopy to optical information processing \cite{rev1}. Bi-directionally pumped resonator have recently became an active sub-area in this field 
\cite{gaeta,vahala,tobias}. Rich symmetry breaking phenomenology \cite{pascal1,pascal3,gl1,chirality} and applications in velocity measurements \cite{vahala2,vahala3,matsko3} have been some of the underpinning driving forces.  A key property of microresonators is their high finesse which can provide three to six orders of magnitude boost to the circulating powers relative to the input one. This creates a variety of opportunities for efficient 
nonlinear control of optical signals, and, in particular, the soliton blockade effect is utilising  high sensitivity of the nonlinear response of one of the intracavity fields towards 
changes in the driving frequency of the counterrotating one.

One of the bottlenecks in studies of the bidirectional microresonators has 
been the absence of a sufficiently rigorous mathematical model to describe their operation in the multimode regime. A variety of  qualitative models \cite{vahala,tobias} have been used to treat  problems of the
nonlinear crosscoupling between the modes, backscattering, and also of the opposing group velocities of the two waves in different ways, see \cite{skr} for a comparative overview. These issues have been reconciled by demonstrating an equivalence of the coupled-mode model
derived from the ab-initio  Maxwell system to a certain form of the envelope equations  \cite{skr,lobanov,cole}. Below, we are applying this methodology  to study the soliton blockade and  accompanying effects.

Following  Ref. \cite{skr}, we define total electric field, $\cE$, in a microresonator as a superposition of the clockwise, '+', and counterclockwise, '$-$', rotating modes. If  $\psi^\pm_\mu(t)$ are the complex mode amplitudes, $\mu=-N/2+1,\cdots,0,\cdots N/2$ is the relative mode number, $t$ is time, and $\ta$ is angle varying along the resonator circumference, then 
$\cE=\big[e^{iM\ta-i\om_+t}\sum_\mu\psi_\mu^+e^{i\mu(\ta-D_1t)}+c.c.\big]
+\big[
e^{iM\ta+i\om_+t}\sum_\mu\psi_\mu^{-*}e^{i\mu(\ta+D_1t)}+c.c.\big]$.
$D_1/2\pi$ is the repetition rate, i.e.,
free spectral range. $M$ is the absolute mode number corresponding to the resonance frequency $\om_{\mu=0}=\om_0$. The first square bracket in the $\cE$-equation corresponds to the clockwise rotating field,  and the second to the counterclockwise one. 
$\om_+$ is the laser frequency pumping the '+' wave, which is detuned from $\om_0$ by $\delta=\om_0-\om_+$, and by $\ep=\om_+-\om_-$ from the pump frequency
of the '$-$' wave, $\om_-$.

If  
$\psi^\pm(t,\ta)=\sum_\mu\psi^\pm_\mu(t) e^{i\mu\ta}$, then  
coupled mode equations for
$\psi^\pm_\mu$ are equivalent to the following equations for the $\psi^\pm$ envelopes \cite{skr}, 
\begin{linenomath}
\begin{subequations}
	\lab{mod} 
\begin{align}
i\p_t\psi^+&=-\tfrac{1}{2}D_{2}\p_\ta^2\psi^+
\lab{moda}
\\
&
+\delta\psi^+-\gamma|\psi^+|^2\psi^+-2\vg^-\psi^+-i\tfrac{1}{2}\kappa(\psi^+-\cH ),
\nn
\\ 
i\p_{t}\psi^{-}&=-\tfrac{1}{2}D_{2}\p_\ta^2\psi^{-}
\lab{modb}\\
&
+\delta\psi^{-}-\gamma|\psi^{-}|^{2}\psi^{-}-2\vg^+\psi^--i\tfrac{1}{2}\kappa(\psi^{-}-{\cal H}e^{i\ep t}),
\nn\\
\vg^\pm&=\gamma \int_0^{2\pi} |\psi^\pm|^2 
\frac{d\ta}{2\pi}=\gamma\sum_\mu|\psi_{\mu}^{\pm}|^2. \lab{modc}
\end{align}
\end{subequations}
\end{linenomath}
In Eqs. \bref{mod}, the offset between the pump frequencies, $\ep\ne 0$, is the only parameter that breaks the symmetry between the two directions and it is used below 
as the main control parameter. 

\begin{figure*}[h!]
	\centering\includegraphics[width=0.9\textwidth]{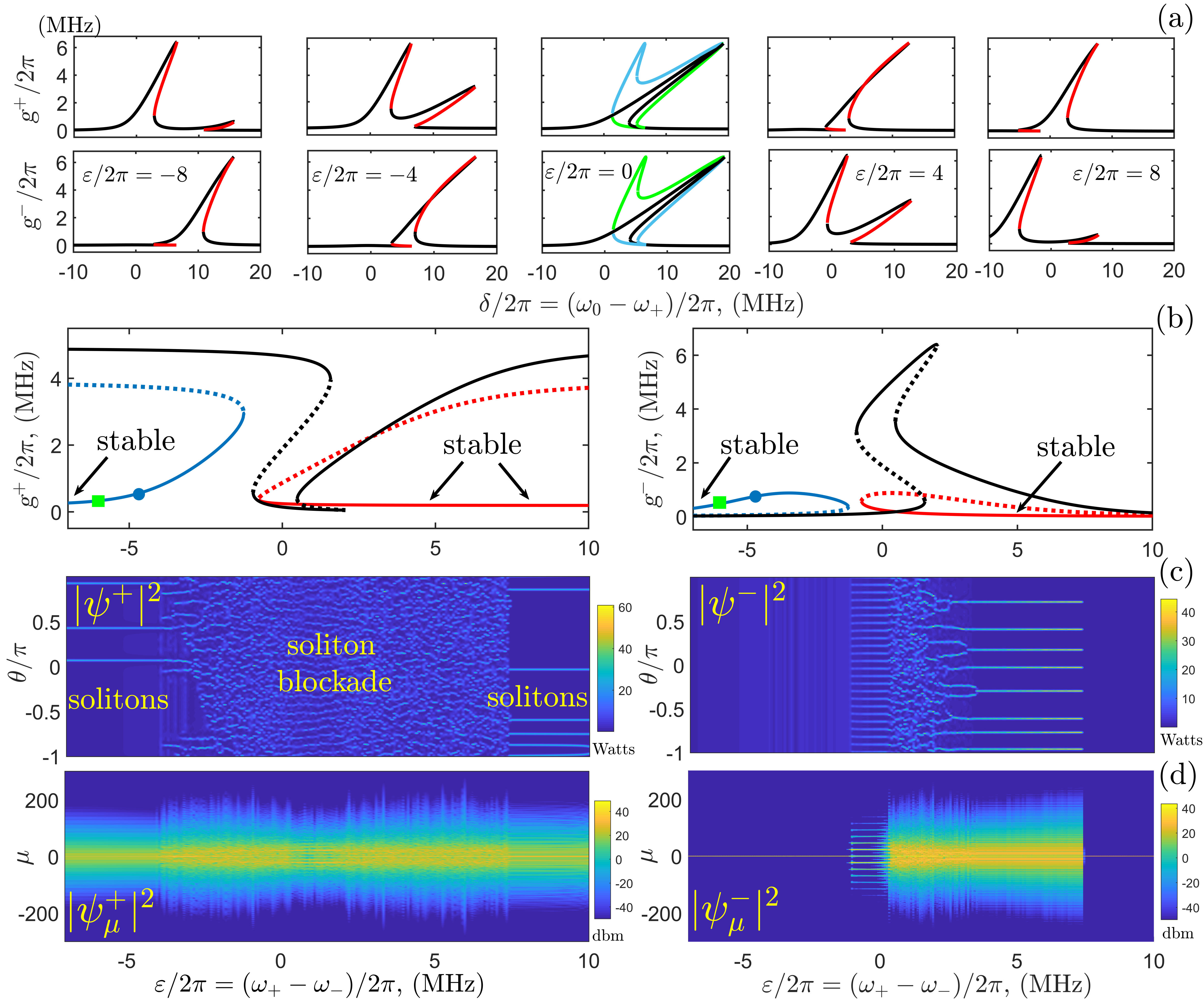}
	\caption{{\Large\bf (a)} Nonlinear resonances of the single mode states, i.e. cw-cw states, $\vg^\pm=\gamma|\wh\psi^\pm|^2$, vs $\delta$ for a set of fixed $\ep$. Same colours show matching branches on the $\vg^+$ and $\vg^-$ plots.   {\Large\bf (b)}~$\vg^\pm$ vs $\ep$ for $\delta$ fixed.  Black lines show the dark-bright resonance around $\ep=0$. Colours and line styles in (b) are independent from (a). Here, they show how branches of the $\vg^+$ (left) and $\vg^-$ (right) cw-cw states match at their turning points.  {\Large\bf (c)}~Soliton blockade effect in the '$+$' field, when  $\ep$ is adiabatically scanned from being large negative, across zero, and towards large positive values. Colour density  shows $|\psi^\pm|^2(\ta)$ along the resonator circumference, $\ta\in [-\pi,\pi)$. Long and sparse straight lines correspond to the soliton-cw states. {\Large\bf (d)} Mode spectrum corresponding to the real space waveforms in (c).  $\cH^2=16$W, $\kappa/2\pi=1.5$MHz, $D_1/2\pi=15$GHz, $\gamma/2\pi=0.4$MHz/W in (a)-(d), $\delta/2\pi=4.5$MHz in (b)-(d), and  $D_2/2\pi=10$kHz in~(c), (d).} 
	\label{f1}
\end{figure*}

\begin{figure*}[h!]
	\centering\includegraphics[width=0.85\textwidth]{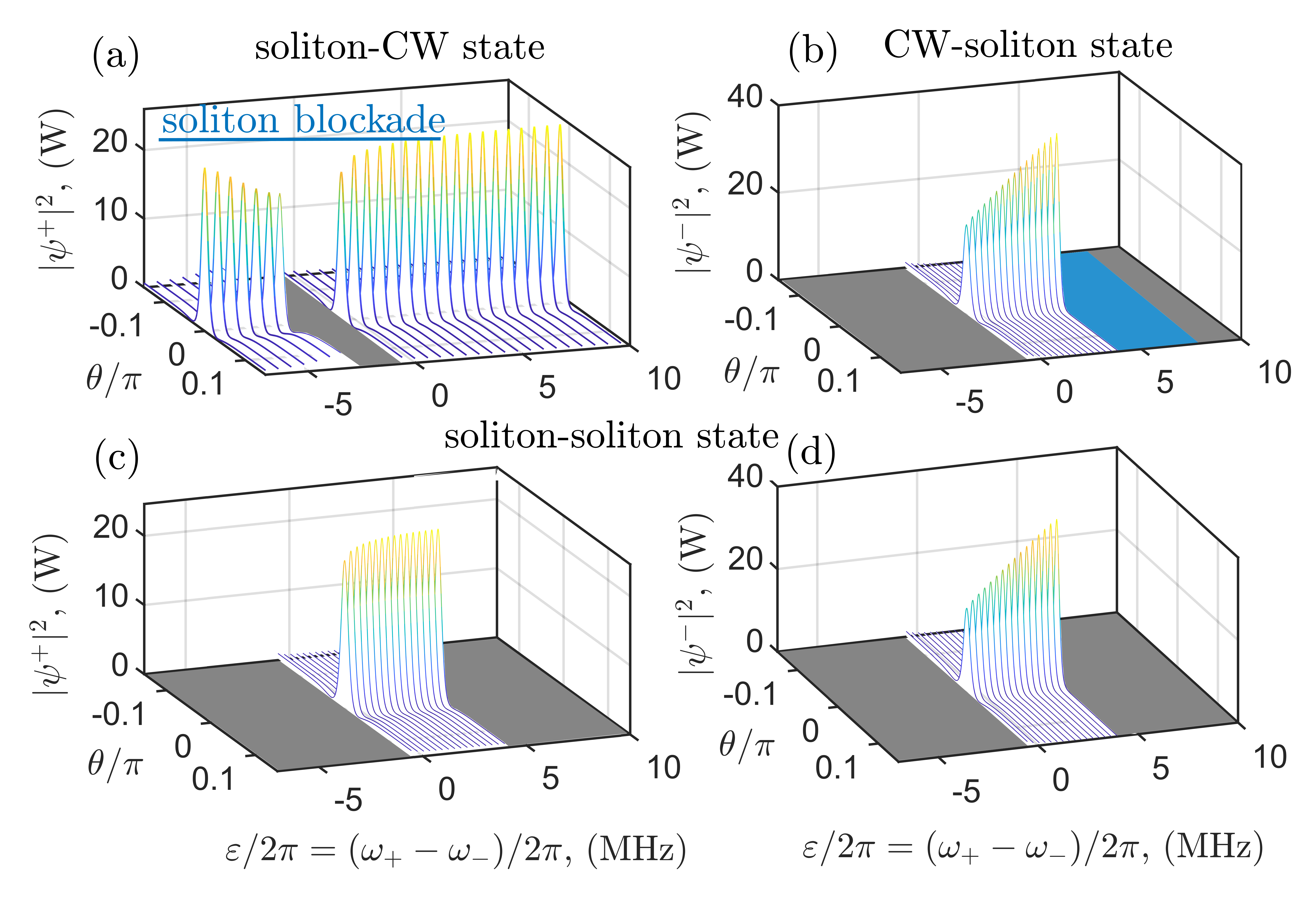}
	\caption{ {\Large\bf (a,b)} Soliton components of the soliton-cw   (a) and of the cw-soliton  (b) states as computed from the time-independent models, Eqs. \bref{sl}, \bref{sl1}, and corresponding to the solitons found in the dynamical simulations, see Fig. \ref{f1}(c). {\Large\bf (c,d)} is the soliton-soliton state. Panel (a) shows the soliton blockade. The blue color in (b) marks the $\ep$ interval where the '$-$' soliton is sustained by the chaotic state in the '$+$' wave, that provides relatively large XPM induced resonance shift, $-2g^+$, compensating for $\ep>0$.} 
	\label{f2}
\end{figure*}

$-2\vg^\pm$ are the negative frequency shifts of the cavity resonance $\om_0$ induced by the nonlinear cross-coupling, i.e., by the XPM (cross-phase modulation) effects. These shifts are different for '$+$' and '$-$' waves.
In the high repetition rate devices, the phase dependent 
four-wave mixing terms inside the XPM part of the Kerr response, $|\psi^\mp|^2\psi^\pm$, oscillate with the fast rate $\sim 2D_1$ \cite{skr,lobanov,cole}.
If finesse $\cF=D_1/\kappa\gg 1$, then the fast oscillations can be disregarded and this is why the XPM  nonlinearity in Eq. \bref{modc}, i.e., $2\vg^\mp\psi^{\pm}$, contains only  powers of the interacting modes and not their phase differences. 
This approximation was recently formally justified and used for Fabry-Perot resonators \cite{cole}, and soon after applied for the ring ones \cite{skr,lobanov}.

$D_{2}$ is the 2nd order dispersion coefficient, so that  the resonator spectrum is $\om_\mu=\om_0+D_1\mu+D_2\mu^2/2!$.
$\gamma$ is the nonlinear parameter \cite{skr}. 
$\kappa$ is the resonance linewidth. 
The intracavity pump  amplitude,
${\cal H}$, is linked to the laser power, $\cal W$, as 
${\cal H}^{2}$ = $\frac{\eta}{\pi} {\cF}  \cal W$, where  $\eta<1$ is the coupling coefficient. 

To study a resonator operating in the single-mode regime in both directions we set 
$\psi^+=\wh\psi^+_{cc}$, $\psi^-=\wh\psi^-_{cc}e^{i\ep t}$, 
$\vg^\pm=\gamma|\wh\psi^\pm_{cc}|^2$, $\p_\ta\wh\psi^\pm_{cc}=\p_t\wh\psi^\pm_{cc}=0$. These are cw-cw states describing 
counterrotating fields with their carrier frequencies  
locked to the respective pump frequencies. In this regime Eqs. \bref{mod} can be reduced to a pair of real equations,
\begin{linenomath}
	\begin{subequations}
		\lab{sl0} 
		\begin{align}
		\gamma\cH^2&=\vg^++4\vg^+\big(\delta-\vg^+-2\vg^-\big)^2/\kappa^2,
		\lab{sl0a}
		\\ 
		\gamma\cH^2&=\vg^-+4\vg^-\big(\delta+\ep-\vg^--2\vg^+\big)^2/\kappa^2.
		\lab{sl0b}
		\end{align}
	\end{subequations}
\end{linenomath}
Large detunings between the two pump frequencies, $|\ep|/\kappa\gg 1$,  effectively decouple nonlinear resonances of the two fields. The latter implies that the $\vg^+$ vs $\delta$ plot shows the nonlinearity tilted resonance originating from $\delta=0$,  and $\vg^-$ has a similar resonance at $\delta=-\ep$. Taking several representative values of $\ep/2\pi$ and other parameters as in  Fig. \ref{f1}(a)  shows that varying $\ep$  from the relatively large negative to large positive values 
drags the $\vg^-$ resonance across the effectively 'immobile' $\vg^+$ one, see Fig. \ref{f1}(a). Two resonances overlap and interact strongly for $|\ep|/\kappa\lesssim 1$. 

For $\ep=0$, Eq. \bref{moda} and Eq. \bref{modb} become symmetric relative
to $\psi^+\to\psi^-$. Hence, there exists a symmetric cw-cw state, $\wh\psi^+_{cc}=\wh\psi^-_{cc}$, see black lines in the $\ep=0$ panels of Fig. \ref{f1}(a). As $\delta$ varies, this state undergoes a symmetry breaking bifurcation into a pair of asymmetric solutions, see blue and green lines in the same panels.
Linear stability  of  $\wh\psi^\pm_{cc}$  relative to noise in the 
$\mu=0$ mode itself and in the $\mu\ne 0$ modes 
has been evaluated using a substitution $\psi^\pm=\big(\wh\psi^\pm_{cc}+\wt\psi^\pm_\mu 
e^{i\mu\ta+\lambda_\mu t}+\wt\psi^{\pm *}_{-\mu} e^{-i\mu\ta+\lambda_\mu^* t}\big)e^{i\ep^\pm t}$, $\ep^+=0$, $\ep^-=\ep$, and linearisation 
of Eqs. \bref{mod}, $|\wh\psi^\pm_{\mu}|/|\wt\psi^\pm_{cc}|\ll 1$.

What is critical for the blockade effect of the '$+$' solitons induced by the '$-$' wave, is  how $\vg^\pm$ in the cw-cw regime vary with $\ep$ for fixed $\delta$, i.e. when the frequency of the '$+$' pump is fixed and frequency of the '$-$' pump is tuned. The respective $\vg^\pm$ vs $\ep$ plots change qualitatively depending on how $\delta$ is chosen. It is critical for  what follows to take $\delta$ within the bistability interval of the $\vg^+$ resonance, e.g., $\delta/2\pi=4.5$MHz as is in Figs. \ref{f1}(b)-(d). This choice of $\delta$ ensures that the '$+$' wave can be excited into a strongly nonlinear state independently if $\ep/\kappa$ is the large positive or large negative. While the '$-$' wave can respond to the pump in a bistable manner only for $|\ep|/\kappa\sim 1$ or less. From this one can anticipate a resonance feature in how the '$-$' wave impacts transmission of the '$+$' wave at $\ep$ around zero.

Three cw-cw states coexisting for $\ep/2\pi \lesssim -7$MHz are well approximated by the $\vg^+$ bistability loop with $\vg^-\approx 0$.
Since dispersion is  anomalous, $D_2>0$, the low amplitude state is stable
and the  high amplitude one is unstable.
$\mu\ne 0$ instabilities of the latter generate solitons in the '$+$'
wave, while the '$-$' wave remains in the cw-state, see Fig. \ref{f1}(c).
The low amplitude state serves as the soliton background.
As $\ep$ is moved towards zero, the  $\vg^-$ power starts building up, and the first effect it brings is that the soliton background becomes unstable relative to the $\mu=0$ noise, see green points in Fig. \ref{f1}(b). This instability  initiates background oscillations which do not make much impact on the soliton cores. Frequency of the oscillations is
the order of $10$ to $10^2$kHz, and they imply that  the counterrotating waves  
are becoming unlocked from  the pump frequencies and tend to lock between themselves.
Soon, the background becomes  unstable relative to the broadband $\mu\ne 0$ perturbations, see blue points in  Fig. \ref{f1}(b). This brings a more detrimental impact on solitons, that is becoming destroyed by the chaotic multi-mode oscillations, see Figs. \ref{f1}(c).

Around $\ep/2\pi\simeq -1$MHz, the low amplitude cw (full blue line) merges with the middle branch (dotted blue line) and both disappear, see Fig. \ref{f1}(b). This implies that the '$+$' solitons become strictly forbidden. Instead,  one can see the formation of the pronounced quasi-stationary roll pattern in the '$-$' component, see Fig. \ref{f1}(c). At the same time, $\vg^+$ corresponding to the high power branch of the cw-cw state starts going down sharply, while the respective $\vg^-$ is rising, which creates the dark-bright resonance,
see black lines in Fig. \ref{f1}(b). Large values of $\vg^-$  
make the net effective detuning of the $\vg^+$ field, i.e., $\delta-2\vg^-$, negative and hence the  '$+$' only solitons impossible. Contrary $\vg^+$ is now relatively small, and as $\ep$ is increased little further the solitons emerge inside the '$-$' wave, see $3\lesssim\ep/2\pi\lesssim 7$MHz in Fig. \ref{f1}(c). For $\ep/2\pi\gtrsim 7$MHz they disappear and the $\vg^-$ resonance again has no impact on the '$+$' one, and therefore the soliton states in the '$+$' field are restored.

The above scenario, see the left panel of Fig. \ref{f1}(c), constitutes the blockade effect for the '$+$' solitons created by tuning the resonance of the '$-$' field across the '$+$' one, see Figs. \ref{f1}(a),(c). 
An approximate analytical expression for the interval of $\ep=\om_+-\om_-$ where the blockade  is taking place can be worked out using following considerations. For $|\ep|/\kappa\gg 1$, $\vg^-\approx 0$ and $\gamma\cH^2\approx\vg^++4\vg^+\left(\delta-\vg^+\right)^2/\kappa^2$. 
Values of $\vg^+$ and $\delta$ at the tip of the nonlinear resonance are connected 
one to the other by $\vg^+=\tfrac{2}{3}\delta+\tfrac{1}{3}\sq{\delta^2-\tfrac{3}{4}\kappa^2}
\approx\delta-\kappa^2/8\delta$ \cite{cole}. Hence $\gamma\cH^2\approx \delta-\kappa^2/16\delta$, and  $\delta$ at the tip is well approximated by 
$\delta_{tip}=\tfrac{1}{2}\gamma\cH^2+\tfrac{1}{2}\sq{\gamma^2\cH^4+\tfrac{1}{4}\kappa^2}$.
Thus, $|\ep|\lesssim\delta_{tip}$ provides quasi-resonant interaction of the counterrotating waves
and can be used as an estimate for the width of the blockade interval. Using parameters as in  
Fig. \ref{f1}  gives $\delta_{tip}/2\pi\approx 6.4$MHz, and  $-6.4\lesssim \ep/2\pi\lesssim 6.4$MHz
is in the remarkably close agreement with numerical data in Fig. \ref{f1}(c).

Steady-state soliton solutions of Eqs. \bref{mod}  can be divided into three broad categories of the soliton-cw, cw-soliton, and soliton-soliton states. Here, the order of words indicates if the soliton component belongs to the '$+$' or to the '$-$' wave. 
The soliton-cw states, which blockade effect we report here, have been observed directly in the numerical scan data shown Fig. \ref{f1}(c). The two remaining states are complimentary to the blockade effect. The state close to the cw-soliton one appears in the scan with its '+' component distorted by the modulations, cf.,  \cite{prx,sb1}. 

The soliton-cw state is sought  in the form
$\psi^+=\wh\psi^+_{sc}(\ta),
\psi^-=\wh\psi^-_{sc}~e^{i\ep t}$,
and satisfies 
\begin{linenomath}
	\begin{subequations}
		\lab{sl} 
		\begin{align}
		\tfrac{1}{2}D_{2}\p_{\ta}^2\wh\psi^+_{sc}&=\big(
		\delta-\gamma|\wh\psi^+_{sc}|^2-2\vg^-\big)\wh\psi^+_{sc}-\tfrac{i\kappa}{2}\big(\wh\psi^+_{sc}-\cH\big),
		\lab{sla}
		\\ 
		\gamma\cH^2&=\vg^-+4\vg^-\big(\delta+\ep-\vg^--2\vg^+\big)^2/\kappa^2.
		\lab{slb}
		\end{align}
	\end{subequations}
\end{linenomath}
Similarly, the cw-soliton states are $\psi^+=\wh\psi^+_{cs},
\psi^-=\wh\psi^-_{cs}(\ta)~e^{i\ep t}$, and satisfy
\begin{linenomath}
	\begin{subequations}
		\lab{sl1} 
		\begin{align}
			\gamma\cH^2&=\vg^++4\vg^+\big(\delta-\vg^+-2\vg^-\big)^2/\kappa^2,
		\lab{sl1a}
		\\ 
		\tfrac{1}{2}D_{2}\p_{\ta}^2\wh\psi^-_{cs}&=\big(
		\delta+\ep-\gamma|\wh\psi^-_{cs}|^2-2\vg^+\big)\wh\psi^-_{cs}-\tfrac{i\kappa}{2}\big(\wh\psi^-
		_{cs}-\cH\big).
		\lab{sl1b}
		\end{align}
	\end{subequations}
\end{linenomath}
The soliton part of  the steady-state soliton-cw solution  plotted vs $\ep$, see  Fig. \ref{f2}(a),  shows 
the blockade interval. The latter is narrower, than the one in the dynamical scan, because the scan picks the dynamic non-stationary regime for $\ep<0$, and drags it  adiabatically to $\ep>0$, see  Fig. \ref{f1}(c). 

The cw-soliton state in Fig. \ref{f2}(b) and the soliton-soliton state in  Figs. \ref{f2}(c),(d) also exist. The latter are found by jointly solving Eqs. \bref{sla}, \bref{sl1b}, and can be anticipated to emerge from the $\ep=0$ point, when this pair of equations is reduced to the  single-component case. 
A range of existence of the '-' solitons in the dynamical scan is wider than is found from the time-independent Eqs. \bref{sl1}, see the blue interval in Fig. \ref{f2}(b). This is because the  scan,  see Fig. \ref{f1}(c), picks the modulated solution in the '$+$' component. The modulated wave pattern provides larger nonlinear shift $2g^+$, for the '$-$' wave than the exact cw component of the cw-soliton state.  Larger negative $-2g^+$ compensate 
for the growing $\ep>0$ and keep the '$-$' soliton afloat for longer, thereby, extending the blockade range.
All three soliton families in Fig. \ref{f2} have been found to have accessible regimes of stable dynamical evolution, also confirmed by the linear stability analysis.  However, the oscillatory instabilities have also been seen for $|\ep|$ close to zero, which could be a subject of a separate investigation.
Interplays of the soliton-blockade with the power imbalance, backscattering, higher-order dispersions  
and Raman scattering also represent interesting problems.
E.g., the Raman effect could be used as a
natural frequency shifter to the soliton channel \cite{vahala,prx}. 

In summary, we have reported how tuning frequency of one of the two 
counterrotating fields causes the dark-bright nonlinear resonance in the single-mode microresonator operation, which translates to the soliton blockade effect in the multi-mode regime.
The soliton blockade constitutes a disruption of the soliton transmission 
in the signal field by changing the frequency of the control field.

Funding: EU  Horizon 2020 Framework 
Programme (812818, MICROCOMB). 
	
\noindent\textbf{Disclosures:} The authors declare no conflicts of interest.

\end{document}